\documentclass[pra,aps,twocolumn,showpacs,epsf]{revtex4}
\usepackage{graphicx}
\usepackage{epsfig}
\usepackage{epsf}
\usepackage{amssymb}
\usepackage{epstopdf}
\usepackage{amsmath}
\usepackage{amstext}
\usepackage{latexsym}
\usepackage[matrix,frame,arrow]{xypic}

\newcommand{\ket}[1]{|#1\rangle}

\begin{document}

\title{Optimal quantum cloning via spin networks}
\author{ Qing Chen$^{1}$}\email{chenqing@mail.ustc.edu.cn}
\author {Jianhua Cheng$^1$}
\author{ Ke-Lin Wang$^1$}
\author {Jiangfeng Du$^{1,2,3}$}\email{djf@ustc.edu.cn}

\address{$^{1}$Department of Modern Physics, University of Science and Technology of China, Hefei 230026, PR China\\
$^{2}$Hefei National Laboratory for Physical Sciences at Microscale and Department of Modern Physics,
University of Science and Technology of China, Hefei, Anhui 230026, PR China\\
$^{3}$Department of Physics, National University of Singapore, 2 Science Drive 3, Singapore 117542
}
\date{\today}

\begin{abstract}
In this paper we present an approach to quantum cloning via free dynamical evolution of spin networks. By properly
designing the network and the couplings between spins, we show that optimal $1\rightarrow M$ phase covariant cloning can
be achieved without any external control. Especially, when M is an odd number, the optimal phase-covariant cloning can
be achieved without ancillas.
Moreover, we demonstrate that the same framework is capable for optimal $1\rightarrow2$ universal cloning.
\end{abstract}
\maketitle

The no-cloning theorem \cite {1} presents that quantum mechanics prohibits perfect cloning of an arbitrary state, which
is one of the most fundamental differences between classical and quantum information processing. This no-go theorem
plays an important role in the security of quantum cryptography \cite {Gisin1}.
Since ideal replication of information is forbidden, it is then interesting to
discuss how close to ideality one can afford to copy an unknown quantum state, namely the upper bound to the fidelity of
approximate cloning \cite {3,Gisin2,5,6,Fan,Chiara3}.
In a pioneering work of Bu\v{z}ek and Hillery \cite {3}, they proposed an optimal $1\rightarrow2$ universal cloning scheme.
Later, Gisin and
Massar presented the unitary transformation leading to optimal $1\rightarrow M$ universal cloning \cite {Gisin2}.
 Other than universal cloning, Bru{\ss} \emph{et al} proposed state-dependent cloning, where
 partial information of the input state is priorly known \cite {5}.
An interesting example is the phase covariant cloning (PCC) \cite {6}, where the input states can be expressed as
$|\psi\rangle=\frac{1}{\sqrt{2}}(|0\rangle+e^{i\phi}|1\rangle)$, namely equatorial states (in the case of qubits).
Since the input state is confined in a subset of the Bloch sphere, higher optimal fidelity is expected, which has been
demonstrated in relevant papers \cite {Fan, Chiara3}.

Currently, approximate quantum cloning machine has been implemented experimentally within several approaches \cite {9,10,11,12}.
However, most of these proposals are based on quantum logic gates and post-selection methods.
In fact, there are other routes
to implement the required quantum protocols. In previous work, quantum computation for a spin network
based on Heisenberg couplings was reported \cite {13,14,15,16,17,18,19,20,Chiara1,Chiara2}.
For example, with unmodulated Heisenberg chains, high fidelity quantum state
transfer can be achieved \cite {14,15,16,17,18,19}. One attracting feature of this approach is that it does not require time
modulation for the qubits couplings.
Once the initial states and the evolutional hamiltonian is determined, the system can faithfully implement designated
computation task through dynamical evolution. Thus, except the preparation of initial states and the readout of
computation results, the whole computation process does not involve external controlling, which provides relatively
longer decoherence time for the system.
Recently, Chiara \emph{et al} \cite {Chiara1,Chiara2} proposed the implementation of $1\rightarrow M$ and
$N\rightarrow M$ PCC (numerically for several special cases)
within this approach. Nevertheless, in their proposal, the $1\rightarrow M$ PCC can
reach the optimal result only in the case of $M=2$, and for arbitrary $M>2$ the fidelity of their PCC machine is far from the
optimal bound.

In this paper, we show that by properly choosing the initial state of the supplementary qubits,
the XXZ model is capable of implementing optimal $1\rightarrow M$ PCC in a spin star network.
In particular, when $M$ is an odd number, we can realize the optimal PCC without the aid of ancillas.
Finally, we demonstrate that optimal $1\rightarrow2$ universal cloning is also available within this scheme.

\begin{figure}
\begin{center}
\epsfig{file=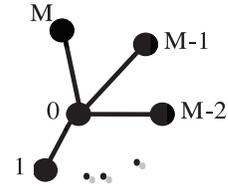,width=0.2\textwidth,height=3cm}
\end{center}
\caption{Spin star network for $1\rightarrow M$ cloning machine.}
\end{figure}

The spin network involved in our scheme forms a star configuration (See Fig. 1). The central qubit is labelled $0$,
 and the outside qubits labelled from $1$ to $M$. The input state is prepared at the central qubit while the outside
 qubits served as supplementary qubits to which the input state will be copied.
 We start with the conventional XXZ Hamiltonian model
\begin{equation}
H=\frac{\cal{J}}{2}\sum_{i=1}^{M}(\sigma_0^x\sigma_{i}^x+\sigma_0^y\sigma_{i}^y+
\lambda\;\sigma_0^z\sigma_{i}^z)+\frac{B}{2}\sum_{i=0}^{M}\sigma_{i}^z,
\end{equation}
where $\sigma^{x,y,z}_i$ are the Pauli matrices corresponding to the i-th spin,
$\cal{J}$ is the exchange coupling between the central site and outer sites, and
$B$ stands for the externally applied magnetic field.
$\lambda$ is the anisotropy parameter which denotes the coupling strength of $z$ direction (when $\lambda=0$, the Hamiltonian
reduces to $XX$ model while $\lambda=1$ it corresponds to Heisenberg model).
Given the Hamiltonian model of Eq. (1), the fidelity is maximized over $B/\cal{J}$, $\lambda$ and $\cal{J}$$t$.
We defined $B^{(M)}$, $\lambda^{(M)}$ and $t^{(M)}$ the values of parameters leading to optimal fidelity. It is helpful
in the later calculation bearing in mind that this Hamiltonian preserves $z$ component of the total angular moment.

In Chiara \emph{et al}'s work \cite {Chiara1,Chiara2}, they addressed PCC within the spin network approach under the
XX model and Heisenberg model. And in their scheme, the initial state of the blank qubits is prepared as $|00..0\rangle$.
Note that the fidelity of PCC within this approach depends both on the Hamiltonian and the initial states.
We hereby discuss this problem in a more general way.
The initial state in our scheme is chosen as
\begin{equation}
|\psi(0)\rangle=|\psi_i\rangle|S(M,k)\rangle, \; \;
|\psi_i\rangle=\alpha|0\rangle+\beta|1\rangle,
\end{equation}
where $\alpha=\cos\frac{\theta}{2}$ , $\beta=e^{i\phi}\sin\frac{\theta}{2}$, $|\psi_i\rangle$ stands for the state to be copied
(prepared at the central qubit) and $|S(M,k)\rangle$  is the initial state of the supplementary spins.
$|S(M,k)\rangle=\frac{1}{\sqrt{C_M^k}}(\hat{P}|\underbrace{000}_k...\underbrace{11}_{M-k} \rangle)$,
where $\hat{P}$ is the total permutation operator ($|0\rangle$ and $|1\rangle$ corresponds to
the eigenstates of $\sigma_z$ with positive and negative eigenvalue respectively).

Let the initial state evolves under the Hamiltonian defined as Eq. (1) for a period of time $t$,
then the output state can be written as (excluding a global phase factor)
\begin{small}
\begin{eqnarray}
|\psi(t)\rangle &=& \alpha\left(f_1(t) |0\rangle |S(M,k)\rangle
+ f_2(t) |1\rangle |S(M,k+1)\rangle \right) \nonumber \\
&+& \beta(g_1(t) |0\rangle |S(M,k-1)\rangle
+ g_2(t) |1\rangle |S(M,k)\rangle),
\end{eqnarray}
\end{small}
where $f_1(t)$, $f_2(t)$, $g_1(t)$, $g_2(t)$ rely on the coefficients $\lambda$, $B$ of the Hamiltonian (1) and
$M$, $k$ of the initial states (2). To get the fidelity of the $1\rightarrow M$ PCC, one needs to
calculate the reduced density matrix of the outside qubits.
For symmetry reasons we only need to calculate one qubit of them. The result is
\begin{widetext}
\begin{small}
\begin{equation}
\rho(t)=\frac{1}{M}
\begin{pmatrix}
|\alpha|^2[k|f_1(t)|^2+ (1+k)|f_2(t)|^2]& \sqrt{k(M-k+1)}\alpha \beta^*f_1(t)g_1(t)^* \\
+ |\beta|^2 [(k-1)|g_1(t)|^2+k |g_2(t)|^2]& + \sqrt{(k+1)(M-k)}\alpha \beta^*f_2(t)g_2(t)^*   \\\\
\sqrt{k(M-k+1)}\alpha^* \beta f_1(t)^*g_1(t) & \;\;\;\;\;\;\;|\alpha|^2[(M-k)|f_1(t)|^2+ (M-k-1))|f_2(t)|^2]\\
+ \sqrt{(k+1)(M-k)}\alpha^* \beta f_2(t)^*g_2(t) & \;\;\;\;\;\;\;+|\beta|^2 [(M-k+1)|g_1(t)|^2+(M-k) |g_2(t)|^2]
\end{pmatrix},
\end{equation}
\end{small}
\end{widetext}

Fidelity $F=\langle\psi_i|\rho(t)|\psi_i\rangle$ is defined to evaluate the performance of the cloner.
For equatorial states ($\theta=\pi/2$), the fidelity can be calculated as
\begin{eqnarray}
F&=&\frac{1}{4}\{2+\frac{\sqrt{k(M-k+1)}}{M}(f_1(t)^*g_1(t)+f_1(t)g_1(t)^*) \nonumber \\
&+&\frac{\sqrt{(M-k)(k+1)}}{M} (f_2(t)^*g_2(t)+f_2(t)g_2(t)^*)  \}.
\end{eqnarray}
Note
\begin{eqnarray}
&&f_1(t)^*g_1(t)+f_1(t)g_1(t)^*+f_2(t)^*g_2(t)+f_2(t)g_2(t)^* \nonumber \\
&\leq& |f_1(t)|^2+|f_2(t)|^2+|g_1(t)|^2+|g_2(t)|^2=2.
\end{eqnarray}
We get
\begin{equation}
F\leq\frac{1}{2}+\frac{1}{2M} \max\{\sqrt{k(M-k+1)},\sqrt{(M-k)(k+1)}\}.
\end{equation}
This equation reveals the important role that initial state of the auxiliary qubits (determined by parameter $k$)
plays in the process of cloning.
It is therefore possible to implement $1\rightarrow M$ PCC with \emph{optimal} fidelity \cite {Fan,Chiara3}
\begin{eqnarray}
&F&=\frac{1}{2}+ \frac{\sqrt{M(M+2)}}{4M}  \quad \textrm{for even} \, M  \nonumber \\
&F&= \frac{1}{2}+ \frac{M+1}{4M}   \quad \textrm{for odd} \, M
\end{eqnarray}
by varying initial states of the supplementary spins.

To work out the exact form of the fidelity shown by Eq. (5), it is convenient to rewrite the Hamiltonian in Eq. (1)
using the Ladder operators
$s^{\pm}_i=(\sigma^x_i\pm i\sigma^y_i)/2$
and
$J^{\pm}=\sum_{outer}s^{\pm}_i$ as
\begin{equation}
H={\cal{J}}(s_0^+ J^- + s_0^- J^+ + 2\lambda\;s_0^z J^z)+B(s_0^z+J^z),
\end{equation}
where $J^z=\sum_{outer}\sigma^{z}_i/2$, $s^z_0=\sigma^z_0/2$.
Consequently, the system can be considered as a resonant interaction between a spin $1/2$ and
a higher spin-$J$ \cite {Hutton}. Such a system is readily analyzed and the eigenstates have the form (For convenience, the
value of coupling strength $\cal{J}$ is set as $1$)
\begin{equation}
|\psi\rangle_{j,m-\frac{1}{2}}^\pm=|0\rangle |j,m-1\rangle + a_{j,m}^\pm|1\rangle |j,m\rangle,
\end{equation}
where
$a_{j,m}^\pm=(\lambda-2m \lambda \pm \sqrt{\lambda^2(2m-1)^2+4 \varepsilon_{j,m}^2}\;)/2\varepsilon_{j,m}$,
$\varepsilon_{j,m}=\sqrt{(j+m)(j-m+1)}$, $j$ is the quantum number associated with eigenstates of
$J^2$ [eigenvalue is $j(j + 1)$], and $m$ is the quantum number for $J^z$.
The corresponding eigenvalues are
\begin{equation}
E_{j,m-\frac{1}{2}}^{\pm}=\{-\lambda+(2m-1)B \pm \sqrt{\lambda^2(2m-1)^2+4\varepsilon_{j,m}^2}\;\}/2.
\end{equation}
The above expressions for eigenstates and eigenvalues do not hold when $m=j+1,-j$. In the case of $m=j+1$,
the eigenstate is $|\psi\rangle_{j,j+\frac{1}{2}}=|0\rangle |j,j\rangle$ and
the corresponding eigenvalue is $E_{j,j+\frac{1}{2}}=j\lambda+(j+\frac{1}{2})B$; while in the case of $m=-j$,
the eigenstate is $|\psi\rangle_{j,-j-\frac{1}{2}}=|1\rangle |j,-j\rangle$ and the corresponding eigenvalue is
$E_{j,-j-\frac{1}{2}}=j\lambda-(j+\frac{1}{2})B$.
The symmetric states can also be expressed as eigenstates of $J^2$ and $J^z$, namely
$\ket{S(M,k)}=\ket{\frac{M}{2},k-\frac{M}{2}}$,
which can be derived in terms of Clebsch-Gordan coefficients \cite {note1}.

Combining the above calculations, we finally reach the required fidelity expression
\begin{equation}
F=\frac{1}{2}+\frac{k(M-k+1)\chi_1(t)-(M-k)(k+1)\chi_2(t)}{M\eta_1 \eta_2},
\end{equation}
where
\begin{eqnarray}
\chi_1(t)&=&\eta_1\cos\frac{\eta_1 t}{2}\sin B t\sin\frac{\eta_2 t}{2} \nonumber \\&-& \lambda(M-2k-1)
\sin\frac{\eta_1 t}{2}\cos B t\sin\frac{\eta_2 t}{2},\nonumber \\
\chi_2(t)&=&\eta_2\cos\frac{\eta_2 t}{2}\sin B t\sin\frac{\eta_1 t}{2}\nonumber \\& -& \lambda(M-2k+1)
\sin\frac{\eta_2 t}{2}\cos B t\sin\frac{\eta_1 t}{2},\nonumber
\end{eqnarray}
and
$\eta_1=\sqrt{4(M-k)(k+1)+(M-2k-1)^2\lambda^2}$,
$\eta_2=\sqrt{4k(M-k+1)+(M-2k+1)^2\lambda^2}$.
\\

$Optimal$ $1\rightarrow M$ $PCC$.
As expected, the optimal $1\rightarrow M$ PCC can be achieved by properly choosing a group of parameters. The
choice that meets the challenge is not unique and here we just provide one applicable solution.
In the case of $M$ is an even number, the initial state of the outside spins is chosen as $|S(M,\frac{M}{2})\rangle$
while the other parameters are chosen as
\begin{eqnarray}
\lambda^{(M)}&=&\sqrt{M(M+2)}; \;\; B^{(M)}=0;\nonumber \\
t^{(M)}&=&\frac{\pi}{\sqrt{2M(M+2)}}.
\end{eqnarray}
For the case of $M$ is an odd number, the initial state is selected as $|S(M,\frac{M-1}{2})\rangle$ and parameters are
\begin{eqnarray}
\lambda^{(M)}&=&\sqrt{\frac{3}{4}(M+1)^2+1}; \;\; B^{(M)}=\frac{M+1}{2};\nonumber \\
t^{(M)}&=&\frac{\pi}{M+1}.
\end{eqnarray}

In the above proposals, the central qubit serves as an ancilla as well as an input port, which makes
the present PCC machines not resource-saving.
When the number of outer spins is an even number ($M=2K$), however, we find that it is possible to make full
use of every qubit. To achieve this goal, with an initial state of supplementary spins prepared as
$|S(M,\frac{M}{2})\rangle$, the system should evolve for a period of $t=\frac{\pi}{\sqrt{2(M+1)(M+2)}}$
without the presence of external magnetic field (coupling strength along $z$ direction is set as $\lambda=M+2$).
In this case, the outcome reduced density matrices of the $2K+1$ qubits are the same and the fidelity
saturates the optimal bound for $1\rightarrow 2K+1$ PCC.
\\

$XX$ and $Heisenberg$ $model$.
In certain experimental systems, the couplings between spins are fixed, it is therefore interesting to calculate the best
performance of the above PCC machine in this circumstance. There are two important Hamiltonian models, known as the
XX model ($\lambda=0$) and the Heisenberg model ($\lambda=1$).
According to Eq. (12), the fidelity under the XX model is
\begin{equation}
F=\frac{1}{2}+\frac{1}{4M}(\gamma_1 \sin \gamma_2 t+ \gamma_2 \sin \gamma_1 t)\sin Bt,
\end{equation}
where $\gamma_1=\sqrt{k(M-k+1)}+\sqrt{(k+1)(M-k)},\; \gamma_2=\sqrt{k(M-k+1)}-\sqrt{(k+1)(M-k)}$.
In this case, the optimal fidelity can be achieved only when $M=2$.
And for $M>2$, the fidelity of our PCC machine can be very close to the optimal bound,
which is well reflected in TABLE. I.
In the case of Heisenberg model, the fidelity can be maximized if there is no external magnetic field and
the evolution time $t=\frac{\pi}{M+1}$. The maximal value is calculated as
\begin{equation}
F=\frac{1}{2}+\frac{1}{M+1}-\frac{2k(M-k)}{M(M+1)^2}.
\end{equation}
Obviously, when the initial state of the outside spins is selected as $|S(M,M)\rangle$ or $|S(M,0)\rangle$,
the maximal fidelity is achieved although it can not meet the optimal bound.

\begin{table}
  \begin{tabular}{|c|c|c|c|c|c|}
\hline
 $M$ &  $\mathcal F_{optimal}$ &$\mathcal F_{max}$&$\cal{J}$$t$&$B/$$\cal{J}$ &$k$\\
\hline
2& 0.853553&0.853553 &3.33216 & 0.471405 & 0\\
3& 0.833333&0.833319 &252.113 &0.0311526 & 1\\
4& 0.806186&0.806131 &108.375 &0.0144940 & 1\\
5& 0.8&0.799642 &27.7507 &0.0566038 &2\\
6& 0.788675&0.788510 &286.127 &0.0274493 &2\\
7& 0.785714&0.785617 &37.3064 &0.0421053 &3\\
8& 0.779508&0.779244 &20.7232 &0.0757989 &3\\
\hline
  \end{tabular}
  \caption{The maximum fidelity $\mathcal F_{max}$ for $1\rightarrow M$ in the case of XX model. $\mathcal F_{optimal}$
is the theoretical optimal $1\rightarrow M$ PCC fidelity. Column 4 and 5 stands for the corresponding evolution time and
external magnetic field strength. The initial state of supplementary qubits is chosen as $S(M,k)$.
 The value $\mathcal F_{max}$ is numerically calculated under the restrictions
$B/$$\cal{J}$ $\in$ $[0.01,1]$ and $\cal{J}$$t$ $\in$ $[0,300]$.
}
  \label{ntom}

\end{table}

It is relatively easier to prepare initial states such as $|S(M,M)\rangle$ and $|S(M,0)\rangle$ when comes to physical
implementations, we hereby make a brief discussion about the above PCC within this case where initial states of
supplementary spins are fixed as $|S(M,M)\rangle$.  One may directly write down the fidelity
\begin{eqnarray}
F&=&\frac{1}{2}+\frac{\cos[(2B+(1+M)\lambda-\sqrt{4M+(M-1)^2\lambda^2}\;)t/2]}{2\sqrt{4M+(M-1)^2\lambda^2}}
\nonumber \\
&-&\frac{\cos[(2B+(1+M)\lambda+\sqrt{4M+(M-1)^2\lambda^2}\;)t/2]}{2\sqrt{4M+(M-1)^2\lambda^2}}.
\end{eqnarray}
by replacing $k=M$ into Eq. (12).
After some straightforward calculations, one arrives at the maximum fidelity
$F=\frac{1}{2}+\frac{1}{2\sqrt{M}}$. A possible solution leading to this maximum value is under
XX model and the corresponding parameters chosen as $t=\frac{\pi}{2\sqrt{M}}$,
$B=\sqrt{M}$. Here our result coincides with Chiara \emph{et al}'s \cite {Chiara1, Chiara2}.
\\

$Optimal$ $1\rightarrow2$ $universal$ $cloning$. An applicable implementation of this protocol is as follows.
With initial state prepared as $(\alpha|0\rangle+\beta|1\rangle)|S(2,1)\rangle$, the system evolves under
Hamiltonian $H=\frac{\cal{J}}{2}\sum_{i=1}^{2}(\sigma_0^x\sigma_{i}^x+\sigma_0^y\sigma_{i}^y+2\sigma_0^z\sigma_{i}^z)$
for $\cal{J}$$t_u=\frac{\pi}{2\sqrt{3}}$. Consequently, the outcome
reduced matrix of the outer two qubits can be calculated as
\begin{equation}
\rho(t_u)=\frac{1}{6}
\begin{pmatrix}
5|\alpha|^2+|\beta|^2& 4\alpha \beta^* \\
4\alpha^* \beta&\;\;\;\;\; |\alpha|^2+5|\beta|^2
\end{pmatrix}.
\end{equation}
Finally, one obtains $F=\frac{5}{6}$, which
is exactly the optimal bound for $1\rightarrow2$ universal cloning.

In summary, we have discussed the implementation of quantum cloning in spin star networks without external
controlling. With XXZ model, we show that initial states of supplementary qubits are crucial for
realizing optimal PCC.
In particular, we provide an applicable choice of the Hamiltonian and the initial state of the auxiliary qubits, in which
case optimal $1\rightarrow M$ PCC is implemented.
Also, we make a brief discussion about the maximal PCC fidelity that XX model and Heisenberg model can afford within our scheme.
Moreover, we demonstrate that optimal $1\rightarrow2$ universal cloning is accessible for this framework.
 Since this scheme does not involve time modulated external controlling, our result
opens up a promising prospect towards robust optimal PCC machines.
Such a prospect is relevant for several experimental systems \cite
{25,26}.

Qing Chen thanks Fei Xu for valuable discussions.
This work is supported by NUS Research Project (Grant No. R-144-000-071-305), the National Fundamental Research
Program (Grant No. 2001CB309300), and National Science
Fund for Distinguished Young Scholars (Grant No.
10425524).


\begin{references}
\bibitem{note1} In the representation as eigenstates of $J^2$ and $J^z$, $\ket{j,m}$ has a degeneracy in
$C_M^r-C_M^{r-1}$ when $j=\frac{1}{2}(M-2r)$, with $r$ ranging from $0$ to $M/2$ if $M$ is even, or $(M-1)/2$
if $M$ is odd \cite {Hutton}. Note that the subspace of $j=\frac{M}{2}$ is non-degenerate.
\end{references}
\end{document}